\documentclass[format=sigconf]{acmart}

\makeatletter
\renewcommand\@formatdoi[1]{\ignorespaces }
\makeatother

\usepackage{hyperref}
\usepackage{booktabs} 
\usepackage{soul} 
\usepackage{graphicx} 
\usepackage{mathtools}
\graphicspath{ {plots/} }
\usepackage{bm} 
\usepackage{multirow}
\usepackage{tikz}
\usepackage{numprint}
\usepackage{todonotes}
\usepackage{subcaption}
\usepackage{balance}
\usepackage{algorithm}
\usepackage{algpseudocode}

\usepackage[autostyle=false, style=english]{csquotes}
\MakeOuterQuote{"}

\newcolumntype{H}{>{\setbox0=\hbox\bgroup}c<{\egroup}@{}} 

\newcommand{\real}{\mathbb{R}}

\newcommand{\normF}[1]{\left\lVert#1\right\rVert^2_F} 

\renewcommand{\vec}[1]{\mathbf{#1}} 

\makeatletter
\newcommand{\mathleft}{\@fleqntrue\@mathmargin1pt}
\newcommand{\mathcenter}{\@fleqnfalse}
\makeatother

\setcopyright{none}
\acmConference[KaRS @RecSys 2018]{KaRS 2018 Workshop on Knowledge-aware and Conversational Recommender Systems} {October 7, 2018}{Vancouver, Canada.} 
\acmDOI{}
\acmISBN{Copyright for the individual papers remains with the authors. Copying permitted for private and academic purposes. This volume is published and copyrighted by its editors} 
\copyrightyear{2018}
\acmPrice{}

\begin{document}
\title{Deriving item features relevance from collaborative domain knowledge}

\author{Maurizio Ferrari Dacrema}
\orcid{0000-0001-7103-2788}
\affiliation{
 \institution{Politecnico di Milano}
}
\email{maurizio.ferrari@polimi.it}

\author{Alberto Gasparin}
\affiliation{
 \institution{Universit\'{a} della Svizzera Italiana}
}
\email{alberto.gasparin@usi.ch}

\author{Paolo Cremonesi}
\orcid{0000-0002-1253-8081}
\affiliation{
 \institution{Politecnico di Milano}
}
\email{paolo.cremonesi@polimi.it}

\renewcommand{\shortauthors}{Ferrari D. et al.}

\begin{abstract}
An Item based recommender system works by computing a similarity between items, which can exploit past user interactions (collaborative filtering) or item features (content based filtering). Collaborative algorithms have been proven to achieve better recommendation quality then content based algorithms in a variety of scenarios, being more effective in modeling user behaviour. However, they can not be applied when items have no interactions at all, i.e. \emph{cold start items}. 
Content based algorithms, which are applicable to cold start items, often require a lot of feature engineering in order to generate useful recommendations. This issue is specifically relevant as the content descriptors become large and heterogeneous.
The focus of this paper is on how to use a collaborative models domain-specific knowledge to build a wrapper feature weighting method which embeds collaborative knowledge in a content based algorithm. 
We present a comparative study for different state of the art algorithms and present a more general model. This machine learning approach to feature weighting shows promising results and high flexibility.
\end{abstract}

\maketitle

\section{Introduction}
Recommender systems aim at guiding the user through the navigation of vast catalogs and in recent years they have become widespread. Among item based algorithms, content based are the most widely used, as they provide good performance and explainability.
Content Based algorithms 
recommend items based on similarities computed via item attributes. Although being applicable in any circumstance in which at least some information about the items is available, they suffer from many drawbacks related to the quality of the item features. It is hard and expensive to provide an accurate and exhaustive description of the item. In recent years the amount of data which is machine-readable on the web has increased substantially, it is therefore possible to build heterogeneous and complex representations for each item (e.g. textual features extracted from web pages). Those representations however often comprise of high number of features and are sparse and noisy, to the point where adding new data hampers the recommendation quality.
Feature weighting, which can be considered as a generalization of feature selection, is a useful tool to improve content based algorithms. Traditional Information Retrieval methods like TF-IDF and BM25 \citep{robertson2004BM25}, while often leading to accuracy improvements, cannot take into account how important are those features from the user point of view.
Collaborative Filtering 
on the other hand determines items similarity by taking into account the user interactions. It is known that collaborative filtering generally outperforms content based filtering even when few ratings for each user are available \cite{CFwinCBF}. 
The main disadvantage of collaborative systems is their inability to compute predictions for new items or new users due to the lack of interactions, this problem is referred to as \emph{cold-start} item. In cold start scenarios only content based algorithm are applicable, this is the case in which improving item description would be most beneficial, and is therefore the main focus of this article. 

The most common approach to tackle the cold-start item problem is to rely on content-based algorithms, whose accuracy is sometimes much poorer.
In this paper we further investigate the cold-start item problem focusing on how to learn feature weights able to better represent feature importance from the user point of view. We provide a comparative study of state of the art algorithms and present a more general model demonstrating its applicability on a wide range of collaborative models.
Moreover we describe a two step approach that:
\begin{enumerate}
\item exploits the capability of a generic collaborative algorithms to model domain-specific user behaviour and achieve state-of-the-art performance for warm items
\item embeds the collaborative knowledge into feature weights
\end{enumerate}

The rest of the paper is organized as follows. In Section \ref{sec:related} we briefly review the literature in the cold-start recommendation domain, in Section \ref{sec:model} our framework is presented and a comparison of the different algorithms is discussed in Section \ref{sec:eval}. Finally conclusions and future works are highlighted in Section \ref{sec:conclusions}

\section{Related Works}\label{sec:related}
Various tools are at our disposal to assess the relevance of a feature. We can distinguish feature weighting algorithms in three categories: filtering, embedding and wrappers \cite{guyon2003introduction}.

Filtering methods usually rely on information retrieval. Methods like TF-IDF or BM25 are not optimized with respect to a predictive model, therefore the resulting weights are not domain-specific and can not take into account the rich collaborative information, even when available.

Embedding methods learn feature weights as a part of the model training, examples of this are UFSM \cite{Elbadrawy:2015} and FBSM \cite{sharma2015FBSM}. Among embedded methods main drawbacks are a complex training phase and noise sensitivity due to the strong coupling of features and interactions.
\emph{User-Specific Feature-based Similarity Models} (UFSM) learns a personalized linear combination of similarity functions known as global similarity functions and can be considered as a special case of Factorization Machines.
\emph{Factorized Bilinear Similarity Models} (FBSM) was proposed as an evolution of UFSM and aims to discover relations among item features. The FBSM similarity matrix is computed as follows:
\begin{equation}
    sim(i,j) = \vec{f}_i^T \vec{W} \vec{f}_j
\label{eq:sim_FBSM}
\end{equation}
where $f$ is the feature vector of the item, $W$ is the matrix of parameters whose diagonal elements represents how well a feature of item $i$ interacts with the same feature of item $j$, while the off diagonal elements determines the correlation among different features. In order to reduce the number of parameters $W$ is represented as the summation of diagonal weights and a low rank approximation of the off-diagonal values:
\begin{equation}
    \vec{W} = \vec{D} + \vec{V}^T \vec{V}
\label{eq:W_FBSM}
\end{equation}
where $D$ is a diagonal matrix having as dimension the number of features, $n_F$, and $V \in \real ^ {n_L \times n_F}$. The number of latent factors $n_L$ is treated as a parameter.

Wrapper methods rely on a two step approach, by learning feature weights on top of an already available model, an example of this is \emph{Least-square Feature Weights} (LFW) \cite{cella2017LFW}. LFW learns feature weights from a SLIM similarity matrix using a simpler model with respect to FBSM.
\begin{equation}
	sim(i,j) = \vec{f}_i^T \vec{D} \vec{f}_j
\end{equation}

All these algorithms, to our best knowledge, have never been subject to a comparative study.

\section{SIMILARITY BASED FEATURE WEIGHTING}\label{sec:model}

Feature weighting can be formulated as a minimization problem whose objective function is:

\begin{equation}
\begin{aligned}
& \operatornamewithlimits{argmin}\limits_{\mathbf{W}}
& & \normF{\vec{S^{(CF)}} - \vec{S^{(W)}}} + \lambda \normF{\vec{D}} + \beta \normF{\vec{V}} \\
\end{aligned}
\label{eq:CBFW_argmin_problem}
\end{equation}
where $\vec{S^{(CF)}}$ is any item-item collaborative similarity, $\vec{S^{(W)}}$ is the similarity function described in Equation \eqref{eq:sim_FBSM}, $\vec{W}$ is the feature weight matrix which captures the relationships between items features and
$\beta$ and $\lambda$ are the regularization terms, we call this model \emph{Collaborative boosted Feature Weighting} (CFW).
This model can either use the latent factors (CFW D+V), as in FBSM, or not (CFW D), as in LFW.

The advantages of learning from a similarity matrix, instead of using the user interactions, are several:
\begin{itemize}
\item High flexibility in choosing the collaborative algorithm which can be treated as a black box
\item Similarity values are less noisy than user interactions
\item The model is simpler and convergence is faster
\end{itemize}

In this paper a two steps hybrid method is presented, in order to easily allow to embed domain-specific user behaviour, as represented by a collaborative model, in a weighted content based recommender. The presented model is easily extendable to other algorithms and domains. The learning phase is composed by two steps.
The goal of the first step is to find the optimal parameters for the collaborative algorithm, to this end a collaborative algorithm is trained and tuned on warm items. The second step applies an embedded method to learn the optimal item feature weights that better approximate the item-item collaborative similarity obtained before.

\begin{table*}[t]
    \begin{tabular}{cl|ccccc|ccccc|}
    \toprule
    &					& \multicolumn{5}{c}{Netflix}	\vline &   \multicolumn{5}{c}{The Movies} \vline\\
    \multicolumn{2}{c}{Algorithm} \vline & Precision & Recall	&	MRR		&	MAP		&	NDCG		& 	Precision & Recall	&	MRR		&	MAP		&	NDCG	\\
    \midrule
    Content&CBF KNN				& 0.0439	&	0.0405	&	0.1177	&	0.0390	&	0.0449		& 0.3885	&	0.0916	&	0.6909	&	0.3166	&	0.1641	\\
    baselines&CBF KNN	IDF			& 0.0439	&	0.0405	&	0.1177	&	0.0390	&	0.0449		& 0.3931	&	0.0930	&	0.6956	&	0.3215	&	0.1662	\\
    &CBF KNN	BM25		& 0.0466	&	0.0410	&	0.1237	&	0.0414	&	0.0462		& 0.3931	&	0.0930	&	0.6956	&	0.3215	&	0.1662	\\
    \midrule
    hybrid feature&FBSM 			& 0.0244	&	0.0240	&	0.0476	&	0.0162	&	0.0199			& 0.2957	& 0.0727	& 0.5503	& 0.2114	& 0.1192	\\
    weights baselines&LFW 			& 0.0679	& 	0.0573	& 	0.1632	& 	0.0631	& 	0.0646			& 0.4135	& 0.0959	& 0.7073	& 0.3442	& 0.1736\\
    \midrule
    &CF KNN 		& 0.0688	& 0.0597	& 0.1585	& 0.0609	& 0.0645		& 0.3891	& 0.0906	& 0.6939	& 0.3192	& 0.1597\\
    &P3alpha 	& \textbf{0.0714}	& \textbf{0.0679}	& \textbf{0.1707}	& \textbf{0.0664}	& \textbf{0.0716}		& 0.3847	& 0.0882	& 0.6911	& 0.3179	& 0.1578\\
    CFW - D&RP3beta 	& 0.0656	& 0.0624	& 0.1643	& 0.0610	& 0.0669		& \textbf{0.4281}	& \textbf{0.1010}	& \textbf{0.7233}	& \textbf{0.3588}	& \textbf{0.1806}\\
    & SLIM RMSE	& 0.0643	& 0.0529	& 0.1572	& 0.0583	& 0.0604		& 0.4058	& 0.0923	& 0.7017	& 0.3372	& 0.1669\\
    &SLIM BPR 	& 0.0685	& 0.0539	& 0.1583	& 0.0618	& 0.0619		& 0.4170	& 0.0967	& 0.7218	& 0.3455	& 0.1723\\
	\bottomrule
   	\end{tabular}
	\caption{Performance of CFW and baselines evaluated on cold items.}
    \label{tab:CFW_and_baselines}
\end{table*}

\subsection{Parameter estimation}
We solve \eqref{eq:CBFW_argmin_problem} via SGD applying Adam \cite{kingma2014adam} which is well suited for problems with noisy and sparse gradients. Note that here our goal is to find weights that will approximate as well as possible the collaborative similarity, this is why we optimize MSE and not BPR, which was used in FBSM. The objective function is therefore:
\begin{equation}
\mathcal{L}_{MSE}(D,V) = \frac{1}{2}\sum\limits_{i \in I}\sum\limits_{j \in I} \left(\hat{s}_{ij} - s_{ij}^{CF}\right)^2 + \lambda \normF{\vec{D}} + \beta \normF{\vec{V}}
\end{equation}

\section{Evaluation}\label{sec:eval}
We performed experiments to confirm that our approach is capable of embedding collaborative knowledge in a content based algorithm improving its recommendation quality in an item cold-start scenario.

\subsection{Dataset}
In order to evaluate our approach we used only item descriptors accessible via web, which we assume will be available for new items, excluding user generated content. The datasets are the following:
\paragraph{\textbf{Netflix}} Enriched with structured and unstructured attributes extracted from IMDB. 
This dataset has 250k users, 6.5k movies, 51k features and 8.8M ratings in 1-5 scale. 
The rating data is enriched with 6.6k binary editorial attributes such as director, actor and genres.
\paragraph{\textbf{The Movies Database}}\footnote{https://www.kaggle.com/rounakbanik/the-movies-dataset}: 45k movies with 190k TMDB editorial features and ratings for 270k users. This dataset has been built from the original one by extracting its 70-cores.

For all the listed datasets, features belonging to less than 5 items or more than 30\% of the items have been removed, as done in \cite{sharma2015FBSM}.
 
\subsection{Evaluation procedure}

\begin{figure}[t]
\centering
\scalebox{0.6}{
\begin{tikzpicture}[every node/.style={minimum size=.3cm-\pgflinewidth, outer sep=0pt}]
	\node[scale=1.5] at (2.75,-0.5) {warm items};
    \draw[draw=black, dashed, ultra thick] (0,0) -- (0,-1);
    \draw[draw=black, dashed, ultra thick] (0,-1) -- (5.5,-1);
    \draw[draw=black, dashed, ultra thick] (5.5,-1) -- (5.5,0);
    
	\node[scale=1.5] at (2,  2.9) {train};
    \node[scale=1.5] at (4.7,2.9) {validation};
	\node[scale=1.5] at (6.2,2.9) {test};
	\node[scale=1.5, rotate=90] at (-0.3,1.3) {users};
    
    \draw[draw=black, ultra thick] (0,0) rectangle (4,2.6);
    \draw[draw=black, ultra thick] (4,0) rectangle (5.5,2.6);
    \draw[draw=black, ultra thick] (5.5,0) rectangle (7,2.6);
    
    \node[minimum size=1.cm] at (2,1.3) {\Huge $A$};
    \node[minimum size=1.cm] at (4.7,1.3) {\Huge $B$};
    \node[minimum size=1.cm] at (6.2,1.3) {\Huge $C$};
\end{tikzpicture}
}\caption{URM split, A and B contains the warm items while C contains the cold items.}
\label{fig:URM_split}
\end{figure}
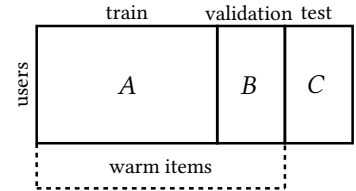

The evaluation procedure consists of two steps.

\paragraph{\textbf{Step 1 - Collaborative algorithm:}} In this step the training of the collaborative algorithm is performed on warm items, which are defined as the union of split A and B, see Figure \ref{fig:URM_split}. Hyper-parameters are tuned with a 20\% holdout validation.
\paragraph{\textbf{Step 2 - Feature weights:}} In this case a cold-item validation is chosen as it better represents the end goal of the algorithm to find weights that perform well on cold items. The collaborative similarity will be learned using only split A with the hyper-parameters found in the previous step, see Figure \ref{fig:URM_split}.
An embedded method is then used to learn the optimal item feature weights that better approximate the item-item collaborative similarity obtained before. The hyper-parameters of the machine learning model are tuned using split B while set C is used for pure testing.

\subsection{Collaborative Similarity}
The collaborative similarity matrices used in our approach are computed using different algorithms: KNN collaborative (using the best-performing similarity among: Cosine, Pearson, Adjusted Cosine, Jaccard), P3alpha \cite{cooper2014P3alpha} (graph based algorithm which models a random walk), Rp3beta \cite{paudel2017Rp3beta} (reranked version of P3alpha), SLIM BPR \cite{ning2011SLIM} and SLIM MSE \cite{levy2013SLIM_ElasticNet}.

\subsection{Results}
Table \ref{tab:CFW_and_baselines} shows the recommendation quality of both pure content based and hybrid baselines, as well as CFW D evaluated on all collaborative similarity models. Table \ref{tab:Netflix_enhanced_CFW} shows the performance of the two components of CFW and FBSM on Netflix, results for the other dataset are omitted as they behave in the same way. 

From Table \ref{tab:CFW_and_baselines} we can see that FBSM performs poorly, which indicates that while it has the power to model complex relations, it is more sensitive to noise and data sparsity than other algorithms. Learning from a similarity matrix, as LFW and CFW D+V does, results in much better results than FBSM. In Table \ref{tab:Netflix_enhanced_CFW} it is possible to see that the latent factor component was able to learn very little, this suggests that while rendering the model more expressive, it introduces noise and numerical instability. Note that the performance using of the diagonal component alone is higher than the one obtained by adding the V component.
However, its effectiveness could be influenced by the feature structure and therefore it might be relevant in some specific cases.
rom Table \ref{tab:CFW_and_baselines} we can see that by using only the diagonal and discarding the latent factor component, the performance improves significantly. 

While LFW only learned feature weights using a SLIM similarity matrix, our results indicate that it is possible to learn from a wide variety of item based algorithms, even those not relying on machine learning. This means that machine learning feature weights can be used on top of already available collaborative algorithm with little effort. Using an intermediate similarity matrix while offering additional degrees of freedom in the selection of the collaborative model, also simplifies the training phase and improves overall performance.

\begin{table}[t]
    \begin{tabular}{clccccc}
    \toprule
    Model	&& Precision & Recall	& MRR		& MAP		& NDCG \\
    \midrule
    	 &D + V 		& 0.0244	& 0.0240	& 0.0476	& 0.0162	& 0.0199 \\
    FBSM &D 		& \textbf{0.0348}	& \textbf{0.0366}	& \textbf{0.0954}	& \textbf{0.0312}	& \textbf{0.0379} \\
      	 &V 		& 0.0138	& 0.0112	& 0.0247	& 0.0071	& 0.0086 \\
    \midrule
    	 &D + V		& 0.0475	& 0.0424	& 0.1336	& 0.0430	& 0.0489 \\
    CFW &D			& \textbf{0.0635}	& \textbf{0.0579}	& \textbf{0.1653}	& \textbf{0.0602}	& \textbf{0.0641} \\
    	 &V			& 0.0412	& 0.0346	& 0.1146	& 0.0335	& 0.0392 \\   
	\bottomrule
   	\end{tabular}
    \caption{Model component contribution on the result of FBSM and CFW on Netflix.}
    \label{tab:Netflix_enhanced_CFW}
\end{table}

\section{Conclusions}
\label{sec:conclusions}
In this paper we presented different state of the art feature weighting methods, compared their performance and proposed a more general framework to effectively apply machine learning feature weighting to boost content based algorithms recommendation quality, embedding user domain-specific behaviour. We also demonstrate high flexibility in the choice of which collaborative algorithm to use.
Future work directions include testing the proposed approach in different datasets and domains, as well as exploring the symmetric problem of using the collaborative similarity to discover item features or to reduce feature noise.

\bibliographystyle{ACM-Reference-Format}
\bibliography{ref_cfwa}
\balance

\end{document}